\documentclass[aps,prd, twocolumn,superscriptaddress,preprintnumbers,floatfix,nofootinbib,notitlepage]{revtex4-1}

\usepackage{graphicx}
\usepackage{amsmath,amsfonts,amssymb}
\usepackage[colorlinks,urlcolor=black,citecolor=blue,linkcolor=black]{hyperref}
\usepackage{color}
\usepackage{verbatim}

\def\lsim{\mathrel{\raise.3ex\hbox{$<$\kern-.75em\lower1ex\hbox{$\sim$}}}}
\def\gsim{\mathrel{\raise.3ex\hbox{$>$\kern-.75em\lower1ex\hbox{$\sim$}}}}

\newcommand{\ie}{{\it i.e.}}
\newcommand{\eg}{{\it e.g.\,}}
\newcommand{\Eq}[1]{Eq.~(\ref{#1})}
\newcommand{\rfn}[1]{(\ref{#1})}

\newcommand{\be}{\begin{equation}}
\newcommand{\ee}{\end{equation}}
\newcommand{\bea}{\begin{equation}\begin{aligned}}
\newcommand{\eea}{\end{aligned}\end{equation}}
\newcommand{\td}{{\rm d}}

\newcommand{\Msun}{M_{\odot}}

\begin{document}

\title{
Cosmological black holes are not described by the Thakurta metric: \\
LIGO-Virgo bounds on PBHs remain unchanged
}

\author{Gert H\"utsi}
\email{gert.hutsi@to.ee}
\affiliation{Laboratory of High Energy and Computational Physics, NICPB, R\"avala pst. 10, 10143 Tallinn, Estonia}

\author{Tomi Koivisto}
\email{timoko@kth.se}
\affiliation{Laboratory of High Energy and Computational Physics, NICPB, R\"avala pst. 10, 10143 Tallinn, Estonia}
\affiliation{Laboratory of Theoretical Physics, University of Tartu, W. Ostwaldi 1, 50411 Tartu, Estonia}

\author{Martti Raidal}
\email{martti.raidal@cern.ch}
\affiliation{Laboratory of High Energy and Computational Physics, NICPB, R\"avala pst. 10, 10143 Tallinn, Estonia}

\author{Ville Vaskonen}
\email{vvaskonen@ifae.es}
\affiliation{Institut de Fisica d'Altes Energies (IFAE), The Barcelona Institute of Science and Technology, Campus UAB, 08193 Bellaterra (Barcelona), Spain}

\author{Hardi Veerm\"ae}
\email{hardi.veermae@cern.ch}
\affiliation{Laboratory of High Energy and Computational Physics, NICPB, R\"avala pst. 10, 10143 Tallinn, Estonia}

\begin{abstract}
We show that the physical conditions which induce the Thakurta metric, recently studied by B{\oe}hm {\it et al.} in the context of time-dependent black hole masses, correspond to a single accreting black hole in the entire Universe filled with isotropic non-interacting dust. In such a case, the physics of black hole accretion is not local but tied to the properties of the whole Universe. We show that radiation, primordial black holes or particle dark matter cannot produce the specific energy flux required for supporting the mass growth of Thakurta black holes. In particular, this solution does not apply to black hole binaries. We conclude that cosmological black holes and their mass growth cannot be described by the Thakurta metric, and thus existing constraints on the primordial black hole abundance from the LIGO-Virgo and the CMB measurements remain valid.

\end{abstract}

\maketitle

\section{Introduction}

If primordial black holes (PBHs) exist, their binary formation starts before the matter-radiation equality. The merger rate of these binaries and the resulting gravitational wave (GW) signal strength both depend on the PHB mass distribution and  abundance~\cite{Nakamura:1997sm,Ioka:1998nz,Raidal:2017mfl,Ali-Haimoud:2017rtz,Raidal:2018bbj,Vaskonen:2019jpv,DeLuca:2020bjf}. In the standard picture, where PBHs formed spatially with a Poisson distribution, and their accretion rate is low, the PBH merger rate today would exceed the one indicated by the LIGO-Virgo observations~\cite{LIGOScientific:2018mvr,Abbott:2020niy} if a large fraction of dark matter (DM) was in $\sim 1-100\Msun$ PBHs. The LIGO-Virgo observations, therefore, constrain the PBH abundance within such mass range~\cite{Raidal:2017mfl,Ali-Haimoud:2017rtz,Raidal:2018bbj,Vaskonen:2019jpv,DeLuca:2020qqa,Hutsi:2020sol}. In addition, the scenario where majority of the observed black hole (BH) mergers originated from PBHs is disfavoured while the presence of a PBH subpopulation within the observed BH mergers remains a viable possibility~\cite{Hall:2020daa,Garcia-Bellido:2020pwq,Hutsi:2020sol,DeLuca:2021wjr,Franciolini:2021tla}.

If the PBH masses change significantly during their evolution, \eg~due to PBHs accretion or mergers, GW signals from PBH binaries are also affected. This phenomenon by itself is simple and robust. However, its realization depends on many nontrivial details such as the nature and properties of DM, initial perturbations and structure formation and the highly non-linear accretion dynamics of BHs. The evolving mass will also affect other early Universe constraints on the PBH abundance, such as all the limits due to the PBH accretion~\cite{Ricotti:2007au,Horowitz:2016lib,Ali-Haimoud:2016mbv,Poulin:2017bwe,Hektor:2018qqw,Serpico:2020ehh,DeLuca:2020fpg}.

In this paper, we consider the possibility that cosmological BHs are described by the Thakurta metric~\cite{1981InJPh..55..304T} \footnote{
We use geometric units $c=G=1$.}
\be \label{g}
    \td s^2 = f(r) \td t^2 - a(t)^2 \left(\frac{\td r^2}{f(r)} + r^2 \td \Omega_2 \right) ,
\ee
where $a$ is the scale factor, $f(r) = 1 - 2m/r $, $H = \dot a/a$ and $m$ is a constant. The Thakurta ansatz was  claimed to imply significant modifications to the constraints on the PBH abundance due to GWs from PBH mergers~\cite{Boehm:2020jwd} and due to energy injection into the CMB due to accretion~\cite{Picker:2021jxl}.

The Thakurta spacetime is a generalized McVittie solution with accretion~\cite{Faraoni:2018xwo}\footnote{
By accretion we mean $\dot M > 0$, which, in general relativity, implies influx of energy.
}. 
The Thakurta metric approximates the FRW metric at $r \gg am$. At $r \ll H^{-1}$, it resembles a Schwarzschild BH, but with an evolving mass: the Misner-Sharp mass enclosed in a sphere of areal radius $R \equiv ar$ is given by
\be\label{eq:mMS}
    M_{\rm MS}(R) = m a + \frac{4\pi}{3} \rho(R) R^3 \,,
\ee
where $\rho(R) \equiv \bar{\rho}/f(R)$ and $\bar \rho$ denote the energy density of the ambient cosmic fluid at $R$ and at spatial infinity. The first term gives the BH mass, while the last term can be interpreted as the mass of matter surrounding the BH. 
In the Newtonian limit, the Misner-Sharp mass coincides with the Newtonian mass at the leading order~\cite{Hayward:1994bu}. Thus, at distances $M \ll R \ll 1/H$, the Thakurta BH is approximately described by a Newtonian point particle with growing mass,
\be\label{ma}
    M = m a \,.
\ee
The accretion rate of such BHs is proportional to the Hubble rate, \ie, $\dot M = H M$, thus tying together cosmological expansion and the local accretion physics of individual BHs.

In general relativity, the Thakurta metric \eqref{g} implies the spherical accretion that the spherical accretion of matter\footnote{In case the radial velocity of matter vanishes exactly, the influx of energy is caused by pure heat flow. Even an infinitesimal flux of matter reverses the direction of the heat flow and the mass growth is dominated by the influx of matter~\cite{Carrera:2009ve}. Some infall of matter, \eg baryons, is expected in any realistic cosmology.} into the BH follows the stress-energy tensor~\cite{Mello:2016irl,Faraoni:2018xwo}
\be
    T^{\mu\nu} = (\rho + P)u^{\mu}u^{\nu} + P g^{\mu\nu} + u^{(\mu}q^{\nu)} ,
\ee
where $u_{\mu}u^{\mu} = 1$, $q_{\mu}u^{\mu}=0$, $q_{\mu}$ describes a radial heat flow with $q_{\mu}q^{\mu}= m^2 H^2/(4\pi r^2 a f)^2$ and $P$ is the pressure of the cosmic fluid. This idealized radial energy flux is completely isotropic and supports the specific accretion of mass given in Eq.~\eqref{ma}. As the Thakurta spacetime is constructed by first {\it postulating} the metric, the properties of the surrounding matter are implied by the Einstein equations and not by any microscopic model of matter. Most obviously, the metric \eqref{g} does not capture the complexities of accretion physics nor the effects that lead to anisotropies and structure formation.

The Thakurta metric includes ambient cosmic fluid that is smooth at the scale of the BH horizon and thus can not correspond to PBHs. Thus, all DM can not be in PBHs, but an additional smooth DM component, \eg, particle or fuzzy DM, is required.
We can naively construct a universe filled with Thakurta PBHs by gluing together several Thakurta spacetimes.\footnote{As there is an energy flow into each patch, additional physical assumptions are needed, \eg, to connect different patches in a way that energy is conserved.} The energy within a spherical volume $V$ is given by the Misner-Sharp mass \eqref{eq:mMS}. If the separation between PBHs is much larger than the horizon, then $f \approx 1$ and thus the ambient energy density is roughly constant, $\rho \approx \bar{\rho}$. Within this approximation, we can consider non-spherical patches with the energy within a patch of volume $V_i$ given by $M_i \approx m_i a + \bar\rho \, V_i$. The average energy density of a universe containing multiple Thakurta patches is then
\be
    \rho_{\rm tot} \approx \frac{\sum_{i} M_i}{\sum_{i} V_i} = \langle m\rangle a\, n_{\rm PBH} + \bar\rho \,,
\ee
where $n_{\rm PBH}$ is the PBH number density, or equivalently the number density of Thakurta patches, and $\langle m\rangle a$ is the average PBH mass. Conservation of PBH number, \ie,  $n_{\rm PBH} \propto a^{-3}$ implies that 
\be\label{eq:a2}
    \rho_{\rm PBH} \approx \langle m\rangle a\, n_{\rm PBH} \propto a^{-2} \,.
\ee
As this scaling is due to the energy transfer from a smooth DM component to the PBHs, it is possible that their combined energy density still scales as $\rho_{\rm tot} \propto a^{-3}$ in which case the DM interpretation would remain valid. 
In this case\footnote{If this mass growth was not due to accretion, \ie, the transfer of energy between different DM components, then the scaling $\rho \propto a^{-2}$ would trivially rule out PBHs as a cold DM candidate which scales as $a^{-3}$.}, it is relevant to understand the physics behind this energy flow.

According to Eq.~\eqref{ma}, the PBH mass growth is quite extreme. For example, between the matter-radiation equality and the redshifts relevant for currently detectable BH mergers, the mass of any PBH would have grown $\sim 3400$ times. It was argued in Ref.~\cite{Boehm:2020jwd} that the mass differences are milder as mass growth should stop once virialized DM halo absorbs the PBH because the Thakurta metric does not describe BHs in such environments. Thus, the postulated effect can be softened further if the PBHs make up most of DM, since then the first virialized structures can form already around matter-radiation equality~\cite{Raidal:2018bbj,Hutsi:2019hlw,Inman:2019wvr,DeLuca:2020jug}. By a similar argument, \Eq{ma} may fail already for as small structures as the PBH binaries or even for BHs that simply move with respect to the cosmic fluid, since such BHs are also not described by the Thakurta metric.

In the following, we argue that the authors of Refs.~\cite{Boehm:2020jwd,Picker:2021jxl} postulated a particular mathematical construction, the metric \rfn{g}, to describe PBHs without checking whether this choice is consistent with our knowledge of cosmology and accretion physics. This allowed them to claim that the constraints on the PBH abundance derived from the LIGO-Virgo observations of BH coalescence are eliminated, opening the possibility that all DM can consist of $30-100~M_\odot$ mass PBHs. This paper aims to point out some phenomenological consequences of the claims made in Ref.~\cite{Boehm:2020jwd,Picker:2021jxl} and to discuss the claimed accretion of PBHs and its effects on the LIGO-Virgo bounds. By studying the consequences of the accretion rate \eqref{ma}, we demonstrate internal inconsistencies of those claims as well as inconsistencies with cosmology, many-body dynamics of PBHs and accretion physics.

\section{Phenomenology}
\label{pheno}

It is not mathematically forbidden to have a configuration of matter around the PBHs that supports the metric \eqref{g}. Whether such a configuration of matter is physically meaningful, what are the needed properties of such a matter, and which are the conditions for accreting it at the rate \eqref{ma} have not been questioned and answered in Ref.~\cite{Boehm:2020jwd}. In the following, we attempt to address some of those questions. We will consider general cosmological scenarios with Thakurta-like accretion. Such scenarios contain the Thakurta PBHs, but are not restricted to it and can correspond to, \eg, perturbations of the exact solution. We conclude that imposing accretion by PBHs, determined by Eq.~\eqref{ma}, 
is unphysical.

We start by considering the case where all DM consists of PBHs, $f_{\rm PBH}=1$, where $f_{\rm PBH}$ is the fraction of DM in PBHs. The central claim and the main result of Ref.~\cite{Boehm:2020jwd} is that, due to the accretion rate \rfn{ma}, $f_{\rm PBH}=1$ is possible for the LIGO BHs in the mass range $30-100~M_\odot$. As there is no particle DM, the PBHs are initially surrounded by the radiation of the Standard Model particles. Around the time of matter-radiation equality, the radiation redshifts and, after the recombination around $z\sim 1100$, the Universe is filled with PBHs and a strong wind of baryons moving with the velocity $v\sim 30~{\rm km/h}$ created by the baryon acoustic oscillations~\cite{Tseliakhovich:2010bj}.

The immediate question arises: in this case, what do the PBHs accrete to grow in mass by several orders of magnitude by the present time? The abundance of baryons is insufficient for that, and, in the early Universe, the baryons move with the velocities of the order $v\sim 30~{\rm km/h}$ which makes them uncapturable by the PBHs. The only possible answer is that the PBHs accrete themselves. We note that this possibility must already deviate from the exact Thakurta metric as PBH DM is certainly not smooth at scales comparable to the BH horizon, and its influx is not described as a heat flow. Regardless, a sufficiently large PHB merger rate needed to support the mass growth of Eq.~\eqref{ma} is impossible for initially Poisson distributed PBHs. This process must, therefore, be described by hierarchical PBH binary mergers whose rate depends on the initial properties of the PBH population, \eg, their mass function or their spatial distribution.  Although the PBH merger rate can be significantly enhanced for initially clustered PBHs,~\cite{Clesse:2016vqa,Raidal:2017mfl,Ballesteros:2018swv,Young:2019gfc,DeLuca:2020jug,Eroshenko:2021oeq} which, by itself, is constrained by the CMB observations~\cite{DeLuca:2021hcf}, it will ultimately be determined by the local parameters shaping the small scale structure of PBHs instead of the specifics of cosmological expansion, which governs Eq.~\eqref{ma}.

In general, PBH mergers cannot grow the average PBH mass by several orders of magnitude. However, even if this was the case, an enormous stochastic GW background would be generated, as each generation of mergers, corresponding to an $\mathcal{O}(2)$ increase of the average PBH mass, converts $\mathcal{O}(5\%)$ of the PBH DM into GWs. In either case, the mass growth asserted by Eq.~\eqref{ma} contradicts studies of PBH structure formation or the GW observations.

Next, consider the case $f_{\rm PBH} < 1$. The neighbourhood of a BH can be modelled using Newtonian physics at distances $M \ll r \ll H^{-1}$. Numerical simulations show that early DM haloes form during the radiation dominated epoch when wider and wider shells of DM surrounding the PBHs decouple from expansion~\cite{Adamek:2019gns}. This results in a density profile that scales as $\rho(r) \propto r^{-9/4}$. Such haloes are stable, but can be disrupted by later close encounters with compact objects, which, when $f_{\rm PBH} \ll 1$, are mostly stars. Again, the relative stability of such DM haloes is in contradiction with the mass growth $m a(t)$. This stability can be understood intuitively by noting that the DM particles will generally not fall into the BH but form a halo due to tidal torque generated by surrounding inhomogeneities. Such orbiting particles are unlikely to fall into the PBH unless there are mechanisms that carry away their energy and angular momentum. The accretion rate \eqref{ma} is therefore not realized.

Taking the accretion rate seriously, one finds that the microscopic theory of the cosmological fluid must have unrealistic properties also for other backgrounds. For example, in a vacuum energy dominated universe, the growth of a Thakurta BH must due to a heat flow carried by vacuum energy. The energy density of the source for the Thakurta metric $\rho(R) \equiv \bar{\rho}/f(R)$ grows towards the BH and diverges at $m=r$. Thus, in a radiation dominated universe in which $T \propto \rho^{1/4}$, the exact Thakurta solution requires an inflow of heat against a temperature gradient in contradiction with the second law of thermodynamics. 

Finally, ignoring the problems with physics behind the mass growth \eqref{ma}, we find (see the Appendix) that the radius of BH binaries with an evolving mass \eqref{ma} shrinks as  $r \propto a^{-3}$. This is a significant modification to binary evolution that can enhance the merger rate. Thus the LIGO-Virgo constraints on PBHs, especially from the stochastic GW background, are not obviously avoided. We must stress again, however, that the Thakurta metric does not describe a binary in the first place and, moreover, accretion into BH binaries can be significantly more involved than accretion into a single BH (see \eg~\cite{Farris:2013uqa,Duez:2018jaf,Munoz:2018tnj}).

\section{Accretion physics}

The radical departure of the BH mass growth $\dot{M} = H M$ from the conventional wisdom $\dot{M} \approx 0$ is due to the accretion of the surrounding energy into the BH, as described by the metric \eqref{g}. Two main approaches have been followed in the studies of the highly nontrivial problem of realistically embedding BHs into the cosmological background, mathematically exact solutions {\'a} la McVittie and physical approximations {\'a} la Zel'dovic.

The McVittie metric~\cite{10.1093/mnras/93.5.325} is an exact solution reducing to the Schwarzschild BH at small radii and to the FLRW at large. It does not, however, allow for accretion~\cite{Faraoni:2018xwo}, and indeed, as most recently shown in Appendix B of Ref.~\cite{DeLuca:2020jug}, the use of the exact McVittie Ansatz leads to the conclusion that the evolution of the BH mass $M$ can be neglected for practical purposes. This supports the generally accepted view that BHs as local systems are decoupled from the global cosmological expansion, but it does not address the issue of accretion. The generalised McVittie metric of the form \eqref{g} seems to describe an accreting BH, but at the price of introducing an imperfect matter source~\cite{Faraoni:2007es,Faraoni:2008tx}. Namely, a radial energy flow in the surrounding fluid is required to incorporate the effect of accretion. It was initially thought that with this amendment, one could alleviate the unphysical properties of the original McVittie solutions, which exhibit spacelike singularities at the horizon and divergent pressures in the matter sector. However, that has been corrected by later calculations, which clarify that the Thakurta geometry does not avoid the unphysical singularities \cite{Carrera:2009ve,Mello:2016irl}. The Thakurta metric with $H \neq 0$ does not describe a BH, but an inhomogeneous expanding universe \cite{Carrera:2009ve,Mello:2016irl}.
In this light, it is difficult to justify the proposal of Refs.~\cite{Boehm:2020jwd,Picker:2021jxl} that \eqref{g} would more realistically describe cosmological BHs.

When considering perfect fluid sources, PBH accretion can be studied approximately. Bondi had considered accretion in a spherically symmetrical star in a steady-state Universe and found it to be proportional to the square of the mass of the star~\cite{10.1093/mnras/112.2.195}. Zel'dovic and Novikov adapted the result to an expanding Universe by arguing that the system is decoupled from the cosmological expansion~\cite{1967SvA....10..602Z}. The BH eats matter from within the accretion radius that, in the case of radiation, is thrice the Schwarzschild radius. The model was refined by Carr and Hawking~\cite{10.1093/mnras/168.2.399}, and further generalised by, \eg, Babichev {\it et al.} \cite{Babichev:2005py,Babichev:2018ubo},
who arrived at the equation (see Section II of Ref.~\cite{Carr:2010wk} for more references and discussion)
\be\label{eq:dotM}
    \dot{M} = 4\pi \frac{\gamma(3\gamma-2)^\frac{3\gamma-2}{2(\gamma-1)}}{4(\gamma-1)^\frac{3}{2}}M^2\rho\,,
\ee
where $\gamma$ is the barotropic index of the perfect fluid and $\rho$ is its cosmological energy density. For dust $\gamma \rightarrow 1$, the accretion radius would be infinite, and for de Sitter $\gamma \rightarrow 0$, so there is no accretion. Thus the qualitative behaviour of Eq.~\eqref{eq:dotM} is reasonable in those limits. The most relevant case is radiation $\gamma=4/3$, for which we can easily integrate the equation to obtain
\be
    \frac{M}{M_0} = \left[ 1 - \frac{9\sqrt{3} M_0}{2t_0}\left( 1- \frac{t}{t_0}\right)\right]^{-1}\,, 
\ee 
where $M_0$ is the mass of the BH at its formation time $t_0$. If the BH is much smaller than the horizon, $M_0 \ll t_0$, this implies negligible accretion. However, the accretion becomes quite significant for PBHs of size comparable to the horizon. In fact, the (special and fine-tuned) self-similar solution yields $M \sim a^2 m$. These special solutions may not be physical, as was argued by Carr and Hawking by taking into account the cosmic expansion (which seems reasonable once $M_0 \sim t_0$). Further, hydrodynamical simulations suggest that a horizon-sized PBH in a perturbed positive-pressure fluid shrinks in relation to the horizon. In conclusion, theoretical justifications to ignore the mass growth of PBHs due to accretion appears fairly robust, also in the Bondi-Zel'dovic-Novikov-based approach.

Nevertheless, in Section \ref{pheno} and Appendix \ref{binary} we considered the implications of the assumption that $M = am$, concluding that such an assumption would not be phenomenologically viable.

\section{Conclusions}

We conclude that the Thakurta metric does not describe realistic cosmological black holes.  Most importantly, the case $f_{\rm PBH} = 1$ when all the DM is in the form of PBHs, which was the main aim of Ref.~\cite{Boehm:2020jwd} to postulate the Thakurta metric, is intrinsically and phenomenologically inconsistent.  Therefore, the existing bounds on the PBH abundance arising from the LIGO-Virgo GW observations and from the BH accretion physics, including the  CMB measurements, remain valid. 

\emph{Note added:} After the initial version of this paper, a critical comment appeared in Ref.~\cite{Boehm:2021kzq}. We replied to their criticism in Ref.~\cite{Hutsi:2021vha}. Around the same time, another paper appeared arguing that the Thakurta metric does not describe a cosmological black hole~\cite{Harada:2021xze}.

\vskip 0.5cm
\noindent
\emph{Acknowledgments.} This work was supported by the Estonian Research Council grants PRG356, PRG803, MOBTB135,  MOBJD381, MOBTT86 and MOBTT5, and by the EU through the European Regional Development Fund CoE program TK133 ``The Dark Side of the Universe." This work was also supported by the grants FPA2017-88915-P and SEV-2016-0588. IFAE is partially funded by the CERCA program of the Generalitat de Catalunya.

\appendix

\section{Binary dynamics with growing mass}
\label{binary}

To address the GW constraints on PBHs, we must consider the dynamics of PBH binaries, which will be modified by the growing mass. For sufficiently large separations, we can apply the Newtonian approximation and study the corresponding two-body problem. Although the Thakurta metric does not describe BHs that move with respect to the cosmic fluid, we will adopt the approach of Ref.~\cite{Boehm:2020jwd} and assume that the mass growth \eqref{ma} also applies to binaries. Neglecting the dynamics of surrounding matter, the Newtonian action for a Thakurta PBH binary\footnote{As the masses of both PBHs in the binary grow, neither of the PBHs is a valid test particle. Therefore the geodesic equation does not apply to the dynamics of the binary. 
},
\bea\label{S}
    S 
    = \int \td t \, a\mu\bigg[ 
        \frac{1}{2}\dot{\mathbf{r}}^2 
    +   \frac{\ddot a}{2a}\mathbf{r}^2 
    +   \frac{a m_{12} }{r} 
    -   \frac{1}{2}\mathbf{r} \cdot \mathbf{T} \cdot \mathbf{r} \bigg],
\eea
can be derived along the lines of~\cite{Ioka:1998nz,Raidal:2018bbj} but by using the evolving mass \eqref{ma} instead of a constant one. Above, $r\equiv|\mathbf{r}|$ is the physical distance, $m_{12} \equiv m_1 + m_2$ and $\mu \equiv m_1m_2/(m_1+m_2)$, and $\mathbf{T}$ describes the tidal forces due to inhomogeneities in the surrounding matter that set the initial angular momentum of the binary.

\begin{figure}
    \centering
    \includegraphics[width=0.45\textwidth]{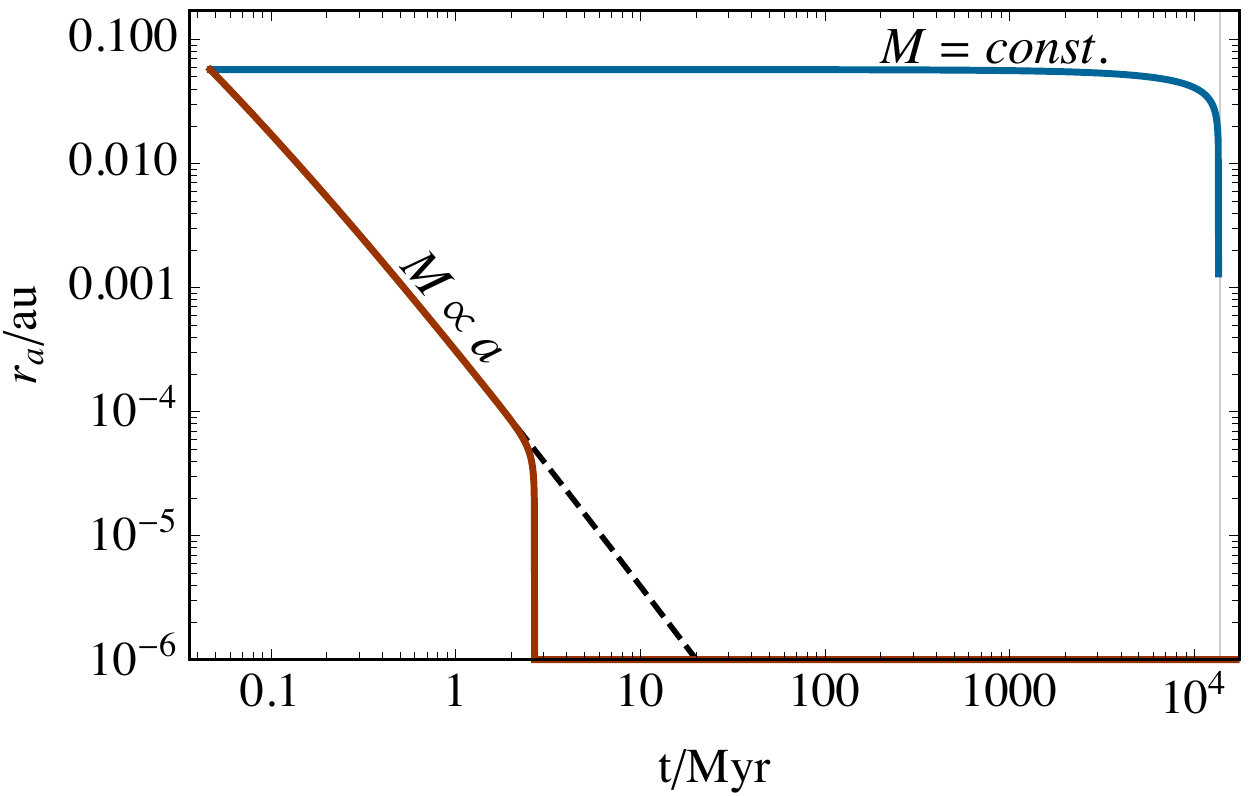}
    \caption{The evolution of circular BH binary with $M \propto a$ (red) and $M = const.$ (blue) from matter-radiation equality to the present day. $r_a = 0.057 \rm au$ at matter radiation equality and the present mass of the binary is $10 \Msun$. The dashed line is proportional to $a^{-3}$.}
    \label{fig:rt}
\end{figure}

Consider the time evolution of a binary that has already formed. In this case, we can neglect the tidal forces and the Hubble flow, that is, the second and the fourth term in \eqref{S}. Conservation of angular momentum is then implied by spherical symmetry. The energy of the binary $E = T + V$ can be split into the kinetic and potential energy, which depend on the scale factor as $T \propto a$ and $V \propto a^2$. In Lagrangian mechanics, time dependence of the energy is by the explicit time-dependence of the Lagrangian, \ie, $\dot E = -\partial_t L = H (-T+2V)$. If the orbital period is much shorter than the Hubble time, as expected for a decoupled binary, then the energy is approximately conserved within a single oscillation. Thus, by the virial theorem, $T = -V/2$. This gives $\dot E = 5H E$. As $E = -a^2 m_1 m_2/r_a$, with $r_a$ the semimajor axis, we obtain that the binary shrinks as
\be\label{ra}
    r_a \propto a^{-3}.
\ee

This behaviour can be generalized to the evolution of the size of any virialized system comprising PBHs obeying $M \propto a$. The argument follows along the same lines as above: the relations $T \propto a$ and $V \propto a^2$ and $\langle T \rangle  = - \langle V\rangle/2$ imply $\dot E = 5H E$. Since the gravitational potential is inversely proportional to the size of the system, $V \propto R^{-1}$, we obtain $R \propto a^{-3}$. Thus, bound systems of PBHs, which can form already at the onset of matter-radiation equality, would shrink approximately as $R \propto a^{-3}$ if the constituent PBHs accrete according to $M \propto a$.

The coalescence time of constant mass binaries scales as $\tau \propto r_a^4$. Therefore, Eq.~\eqref{ra} translates into a rapid decrease of the coalescence time $\tau \propto a^{-12}$, and in the scenario proposed in Ref.~\cite{Boehm:2020jwd}, binaries will shrink due to mass growth until GW emission takes over, after which the binary merges rapidly. For illustration, Fig.~\ref{fig:rt} shows the evolution of the semimajor axis of a circular~\footnote{Eccentric binaries will merge faster than circular ones.} binary with constant mass and with $M \propto a$. The latter evolves as $\dot r_a = - 3Hr_a - (16/5)M(t)^3 r_a^{-3}$, where the last term comes from GW emission~\cite{Peters:1964zz}. Therefore, the mass growth postulated in Eq.~\eqref{ma} can enhance the GW emission from PBH binaries instead of suppressing it. Moreover, this enhancement can be sufficiently strong to force most binaries to merge by the present time, thus reducing the PBH merger rate at low redshifts. In this case, constraints on the PBH abundance arise from the non-observation of a stochastic GW background.

\bibliography{PBH}

\begin{thebibliography}{55}%
\makeatletter
\providecommand \@ifxundefined [1]{%
 \@ifx{#1\undefined}
}%
\providecommand \@ifnum [1]{%
 \ifnum #1\expandafter \@firstoftwo
 \else \expandafter \@secondoftwo
 \fi
}%
\providecommand \@ifx [1]{%
 \ifx #1\expandafter \@firstoftwo
 \else \expandafter \@secondoftwo
 \fi
}%
\providecommand \natexlab [1]{#1}%
\providecommand \enquote  [1]{``#1''}%
\providecommand \bibnamefont  [1]{#1}%
\providecommand \bibfnamefont [1]{#1}%
\providecommand \citenamefont [1]{#1}%
\providecommand \href@noop [0]{\@secondoftwo}%
\providecommand \href [0]{\begingroup \@sanitize@url \@href}%
\providecommand \@href[1]{\@@startlink{#1}\@@href}%
\providecommand \@@href[1]{\endgroup#1\@@endlink}%
\providecommand \@sanitize@url [0]{\catcode `\\12\catcode `\$12\catcode
  `\&12\catcode `\#12\catcode `\^12\catcode `\_12\catcode `\%12\relax}%
\providecommand \@@startlink[1]{}%
\providecommand \@@endlink[0]{}%
\providecommand \url  [0]{\begingroup\@sanitize@url \@url }%
\providecommand \@url [1]{\endgroup\@href {#1}{\urlprefix }}%
\providecommand \urlprefix  [0]{URL }%
\providecommand \Eprint [0]{\href }%
\providecommand \doibase [0]{http://dx.doi.org/}%
\providecommand \selectlanguage [0]{\@gobble}%
\providecommand \bibinfo  [0]{\@secondoftwo}%
\providecommand \bibfield  [0]{\@secondoftwo}%
\providecommand \translation [1]{[#1]}%
\providecommand \BibitemOpen [0]{}%
\providecommand \bibitemStop [0]{}%
\providecommand \bibitemNoStop [0]{.\EOS\space}%
\providecommand \EOS [0]{\spacefactor3000\relax}%
\providecommand \BibitemShut  [1]{\csname bibitem#1\endcsname}%
\let\auto@bib@innerbib\@empty
\bibitem [{\citenamefont {Nakamura}\ \emph {et~al.}(1997)\citenamefont
  {Nakamura}, \citenamefont {Sasaki}, \citenamefont {Tanaka},\ and\
  \citenamefont {Thorne}}]{Nakamura:1997sm}%
  \BibitemOpen
  \bibfield  {author} {\bibinfo {author} {\bibfnamefont {T.}~\bibnamefont
  {Nakamura}}, \bibinfo {author} {\bibfnamefont {M.}~\bibnamefont {Sasaki}},
  \bibinfo {author} {\bibfnamefont {T.}~\bibnamefont {Tanaka}}, \ and\ \bibinfo
  {author} {\bibfnamefont {K.~S.}\ \bibnamefont {Thorne}},\ }\href {\doibase
  10.1086/310886} {\bibfield  {journal} {\bibinfo  {journal} {Astrophys. J.
  Lett.}\ }\textbf {\bibinfo {volume} {487}},\ \bibinfo {pages} {L139}
  (\bibinfo {year} {1997})},\ \Eprint {http://arxiv.org/abs/astro-ph/9708060}
  {arXiv:astro-ph/9708060} \BibitemShut {NoStop}%
\bibitem [{\citenamefont {Ioka}\ \emph {et~al.}(1998)\citenamefont {Ioka},
  \citenamefont {Chiba}, \citenamefont {Tanaka},\ and\ \citenamefont
  {Nakamura}}]{Ioka:1998nz}%
  \BibitemOpen
  \bibfield  {author} {\bibinfo {author} {\bibfnamefont {K.}~\bibnamefont
  {Ioka}}, \bibinfo {author} {\bibfnamefont {T.}~\bibnamefont {Chiba}},
  \bibinfo {author} {\bibfnamefont {T.}~\bibnamefont {Tanaka}}, \ and\ \bibinfo
  {author} {\bibfnamefont {T.}~\bibnamefont {Nakamura}},\ }\href {\doibase
  10.1103/PhysRevD.58.063003} {\bibfield  {journal} {\bibinfo  {journal} {Phys.
  Rev. D}\ }\textbf {\bibinfo {volume} {58}},\ \bibinfo {pages} {063003}
  (\bibinfo {year} {1998})},\ \Eprint {http://arxiv.org/abs/astro-ph/9807018}
  {arXiv:astro-ph/9807018} \BibitemShut {NoStop}%
\bibitem [{\citenamefont {Raidal}\ \emph {et~al.}(2017)\citenamefont {Raidal},
  \citenamefont {Vaskonen},\ and\ \citenamefont
  {{{Veerm\"ae}}}}]{Raidal:2017mfl}%
  \BibitemOpen
  \bibfield  {author} {\bibinfo {author} {\bibfnamefont {M.}~\bibnamefont
  {Raidal}}, \bibinfo {author} {\bibfnamefont {V.}~\bibnamefont {Vaskonen}}, \
  and\ \bibinfo {author} {\bibfnamefont {H.}~\bibnamefont {{{Veerm\"ae}}}},\
  }\href {\doibase 10.1088/1475-7516/2017/09/037} {\bibfield  {journal}
  {\bibinfo  {journal} {JCAP}\ }\textbf {\bibinfo {volume} {1709}},\ \bibinfo
  {pages} {037} (\bibinfo {year} {2017})},\ \Eprint
  {http://arxiv.org/abs/1707.01480} {arXiv:1707.01480 [astro-ph.CO]}
  \BibitemShut {NoStop}%
\bibitem [{\citenamefont {Ali-Haimoud}\ \emph {et~al.}(2017)\citenamefont
  {Ali-Haimoud}, \citenamefont {Kovetz},\ and\ \citenamefont
  {Kamionkowski}}]{Ali-Haimoud:2017rtz}%
  \BibitemOpen
  \bibfield  {author} {\bibinfo {author} {\bibfnamefont {Y.}~\bibnamefont
  {Ali-Haimoud}}, \bibinfo {author} {\bibfnamefont {E.~D.}\ \bibnamefont
  {Kovetz}}, \ and\ \bibinfo {author} {\bibfnamefont {M.}~\bibnamefont
  {Kamionkowski}},\ }\href {\doibase 10.1103/PhysRevD.96.123523} {\bibfield
  {journal} {\bibinfo  {journal} {Phys. Rev.}\ }\textbf {\bibinfo {volume}
  {D96}},\ \bibinfo {pages} {123523} (\bibinfo {year} {2017})},\ \Eprint
  {http://arxiv.org/abs/1709.06576} {arXiv:1709.06576 [astro-ph.CO]}
  \BibitemShut {NoStop}%
\bibitem [{\citenamefont {Raidal}\ \emph {et~al.}(2019)\citenamefont {Raidal},
  \citenamefont {Spethmann}, \citenamefont {Vaskonen},\ and\ \citenamefont
  {Veerm\"ae}}]{Raidal:2018bbj}%
  \BibitemOpen
  \bibfield  {author} {\bibinfo {author} {\bibfnamefont {M.}~\bibnamefont
  {Raidal}}, \bibinfo {author} {\bibfnamefont {C.}~\bibnamefont {Spethmann}},
  \bibinfo {author} {\bibfnamefont {V.}~\bibnamefont {Vaskonen}}, \ and\
  \bibinfo {author} {\bibfnamefont {H.}~\bibnamefont {Veerm\"ae}},\ }\href
  {\doibase 10.1088/1475-7516/2019/02/018} {\bibfield  {journal} {\bibinfo
  {journal} {JCAP}\ }\textbf {\bibinfo {volume} {02}},\ \bibinfo {pages} {018}
  (\bibinfo {year} {2019})},\ \Eprint {http://arxiv.org/abs/1812.01930}
  {arXiv:1812.01930 [astro-ph.CO]} \BibitemShut {NoStop}%
\bibitem [{\citenamefont {Vaskonen}\ and\ \citenamefont
  {Veerm\"ae}(2020)}]{Vaskonen:2019jpv}%
  \BibitemOpen
  \bibfield  {author} {\bibinfo {author} {\bibfnamefont {V.}~\bibnamefont
  {Vaskonen}}\ and\ \bibinfo {author} {\bibfnamefont {H.}~\bibnamefont
  {Veerm\"ae}},\ }\href {\doibase 10.1103/PhysRevD.101.043015} {\bibfield
  {journal} {\bibinfo  {journal} {Phys. Rev. D}\ }\textbf {\bibinfo {volume}
  {101}},\ \bibinfo {pages} {043015} (\bibinfo {year} {2020})},\ \Eprint
  {http://arxiv.org/abs/1908.09752} {arXiv:1908.09752 [astro-ph.CO]}
  \BibitemShut {NoStop}%
\bibitem [{\citenamefont {De~Luca}\ \emph
  {et~al.}(2020{\natexlab{a}})\citenamefont {De~Luca}, \citenamefont
  {Franciolini}, \citenamefont {Pani},\ and\ \citenamefont
  {Riotto}}]{DeLuca:2020bjf}%
  \BibitemOpen
  \bibfield  {author} {\bibinfo {author} {\bibfnamefont {V.}~\bibnamefont
  {De~Luca}}, \bibinfo {author} {\bibfnamefont {G.}~\bibnamefont
  {Franciolini}}, \bibinfo {author} {\bibfnamefont {P.}~\bibnamefont {Pani}}, \
  and\ \bibinfo {author} {\bibfnamefont {A.}~\bibnamefont {Riotto}},\ }\href
  {\doibase 10.1088/1475-7516/2020/04/052} {\bibfield  {journal} {\bibinfo
  {journal} {JCAP}\ }\textbf {\bibinfo {volume} {04}},\ \bibinfo {pages} {052}
  (\bibinfo {year} {2020}{\natexlab{a}})},\ \Eprint
  {http://arxiv.org/abs/2003.02778} {arXiv:2003.02778 [astro-ph.CO]}
  \BibitemShut {NoStop}%
\bibitem [{\citenamefont {Abbott}\ \emph {et~al.}(2019)\citenamefont {Abbott}
  \emph {et~al.}}]{LIGOScientific:2018mvr}%
  \BibitemOpen
  \bibfield  {author} {\bibinfo {author} {\bibfnamefont {B.}~\bibnamefont
  {Abbott}} \emph {et~al.} (\bibinfo {collaboration} {LIGO Scientific,
  Virgo}),\ }\href {\doibase 10.1103/PhysRevX.9.031040} {\bibfield  {journal}
  {\bibinfo  {journal} {Phys. Rev. X}\ }\textbf {\bibinfo {volume} {9}},\
  \bibinfo {pages} {031040} (\bibinfo {year} {2019})},\ \Eprint
  {http://arxiv.org/abs/1811.12907} {arXiv:1811.12907 [astro-ph.HE]}
  \BibitemShut {NoStop}%
\bibitem [{\citenamefont {Abbott}\ \emph {et~al.}(2020)\citenamefont {Abbott}
  \emph {et~al.}}]{Abbott:2020niy}%
  \BibitemOpen
  \bibfield  {author} {\bibinfo {author} {\bibfnamefont {R.}~\bibnamefont
  {Abbott}} \emph {et~al.} (\bibinfo {collaboration} {LIGO Scientific,
  Virgo}),\ }\href@noop {} {\  (\bibinfo {year} {2020})},\ \Eprint
  {http://arxiv.org/abs/2010.14527} {arXiv:2010.14527 [gr-qc]} \BibitemShut
  {NoStop}%
\bibitem [{\citenamefont {De~Luca}\ \emph
  {et~al.}(2020{\natexlab{b}})\citenamefont {De~Luca}, \citenamefont
  {Franciolini}, \citenamefont {Pani},\ and\ \citenamefont
  {Riotto}}]{DeLuca:2020qqa}%
  \BibitemOpen
  \bibfield  {author} {\bibinfo {author} {\bibfnamefont {V.}~\bibnamefont
  {De~Luca}}, \bibinfo {author} {\bibfnamefont {G.}~\bibnamefont
  {Franciolini}}, \bibinfo {author} {\bibfnamefont {P.}~\bibnamefont {Pani}}, \
  and\ \bibinfo {author} {\bibfnamefont {A.}~\bibnamefont {Riotto}},\ }\href
  {\doibase 10.1088/1475-7516/2020/06/044} {\bibfield  {journal} {\bibinfo
  {journal} {JCAP}\ }\textbf {\bibinfo {volume} {06}},\ \bibinfo {pages} {044}
  (\bibinfo {year} {2020}{\natexlab{b}})},\ \Eprint
  {http://arxiv.org/abs/2005.05641} {arXiv:2005.05641 [astro-ph.CO]}
  \BibitemShut {NoStop}%
\bibitem [{\citenamefont {H\"utsi}\ \emph
  {et~al.}(2021{\natexlab{a}})\citenamefont {H\"utsi}, \citenamefont {Raidal},
  \citenamefont {Vaskonen},\ and\ \citenamefont {Veerm\"ae}}]{Hutsi:2020sol}%
  \BibitemOpen
  \bibfield  {author} {\bibinfo {author} {\bibfnamefont {G.}~\bibnamefont
  {H\"utsi}}, \bibinfo {author} {\bibfnamefont {M.}~\bibnamefont {Raidal}},
  \bibinfo {author} {\bibfnamefont {V.}~\bibnamefont {Vaskonen}}, \ and\
  \bibinfo {author} {\bibfnamefont {H.}~\bibnamefont {Veerm\"ae}},\ }\href
  {\doibase 10.1088/1475-7516/2021/03/068} {\bibfield  {journal} {\bibinfo
  {journal} {JCAP}\ }\textbf {\bibinfo {volume} {03}},\ \bibinfo {pages} {068}
  (\bibinfo {year} {2021}{\natexlab{a}})},\ \Eprint
  {http://arxiv.org/abs/2012.02786} {arXiv:2012.02786 [astro-ph.CO]}
  \BibitemShut {NoStop}%
\bibitem [{\citenamefont {Hall}\ \emph {et~al.}(2020)\citenamefont {Hall},
  \citenamefont {Gow},\ and\ \citenamefont {Byrnes}}]{Hall:2020daa}%
  \BibitemOpen
  \bibfield  {author} {\bibinfo {author} {\bibfnamefont {A.}~\bibnamefont
  {Hall}}, \bibinfo {author} {\bibfnamefont {A.~D.}\ \bibnamefont {Gow}}, \
  and\ \bibinfo {author} {\bibfnamefont {C.~T.}\ \bibnamefont {Byrnes}},\
  }\href {\doibase 10.1103/PhysRevD.102.123524} {\bibfield  {journal} {\bibinfo
   {journal} {Phys. Rev. D}\ }\textbf {\bibinfo {volume} {102}},\ \bibinfo
  {pages} {123524} (\bibinfo {year} {2020})},\ \Eprint
  {http://arxiv.org/abs/2008.13704} {arXiv:2008.13704 [astro-ph.CO]}
  \BibitemShut {NoStop}%
\bibitem [{\citenamefont {Garc\'\i{}a-Bellido}\ \emph
  {et~al.}(2021)\citenamefont {Garc\'\i{}a-Bellido}, \citenamefont {Nu\~no
  Siles},\ and\ \citenamefont {Ruiz~Morales}}]{Garcia-Bellido:2020pwq}%
  \BibitemOpen
  \bibfield  {author} {\bibinfo {author} {\bibfnamefont {J.}~\bibnamefont
  {Garc\'\i{}a-Bellido}}, \bibinfo {author} {\bibfnamefont {J.~F.}\
  \bibnamefont {Nu\~no Siles}}, \ and\ \bibinfo {author} {\bibfnamefont
  {E.}~\bibnamefont {Ruiz~Morales}},\ }\href {\doibase
  10.1016/j.dark.2021.100791} {\bibfield  {journal} {\bibinfo  {journal} {Phys.
  Dark Univ.}\ }\textbf {\bibinfo {volume} {31}},\ \bibinfo {pages} {100791}
  (\bibinfo {year} {2021})},\ \Eprint {http://arxiv.org/abs/2010.13811}
  {arXiv:2010.13811 [astro-ph.CO]} \BibitemShut {NoStop}%
\bibitem [{\citenamefont {De~Luca}\ \emph
  {et~al.}(2021{\natexlab{a}})\citenamefont {De~Luca}, \citenamefont
  {Franciolini}, \citenamefont {Pani},\ and\ \citenamefont
  {Riotto}}]{DeLuca:2021wjr}%
  \BibitemOpen
  \bibfield  {author} {\bibinfo {author} {\bibfnamefont {V.}~\bibnamefont
  {De~Luca}}, \bibinfo {author} {\bibfnamefont {G.}~\bibnamefont
  {Franciolini}}, \bibinfo {author} {\bibfnamefont {P.}~\bibnamefont {Pani}}, \
  and\ \bibinfo {author} {\bibfnamefont {A.}~\bibnamefont {Riotto}},\ }\href
  {\doibase 10.1088/1475-7516/2021/05/003} {\bibfield  {journal} {\bibinfo
  {journal} {JCAP}\ }\textbf {\bibinfo {volume} {05}},\ \bibinfo {pages} {003}
  (\bibinfo {year} {2021}{\natexlab{a}})},\ \Eprint
  {http://arxiv.org/abs/2102.03809} {arXiv:2102.03809 [astro-ph.CO]}
  \BibitemShut {NoStop}%
\bibitem [{\citenamefont {Franciolini}\ \emph {et~al.}(2021)\citenamefont
  {Franciolini}, \citenamefont {Baibhav}, \citenamefont {De~Luca},
  \citenamefont {Ng}, \citenamefont {Wong}, \citenamefont {Berti},
  \citenamefont {Pani}, \citenamefont {Riotto},\ and\ \citenamefont
  {Vitale}}]{Franciolini:2021tla}%
  \BibitemOpen
  \bibfield  {author} {\bibinfo {author} {\bibfnamefont {G.}~\bibnamefont
  {Franciolini}}, \bibinfo {author} {\bibfnamefont {V.}~\bibnamefont
  {Baibhav}}, \bibinfo {author} {\bibfnamefont {V.}~\bibnamefont {De~Luca}},
  \bibinfo {author} {\bibfnamefont {K.~K.~Y.}\ \bibnamefont {Ng}}, \bibinfo
  {author} {\bibfnamefont {K.~W.~K.}\ \bibnamefont {Wong}}, \bibinfo {author}
  {\bibfnamefont {E.}~\bibnamefont {Berti}}, \bibinfo {author} {\bibfnamefont
  {P.}~\bibnamefont {Pani}}, \bibinfo {author} {\bibfnamefont {A.}~\bibnamefont
  {Riotto}}, \ and\ \bibinfo {author} {\bibfnamefont {S.}~\bibnamefont
  {Vitale}},\ }\href@noop {} {\  (\bibinfo {year} {2021})},\ \Eprint
  {http://arxiv.org/abs/2105.03349} {arXiv:2105.03349 [gr-qc]} \BibitemShut
  {NoStop}%
\bibitem [{\citenamefont {Ricotti}\ \emph {et~al.}(2008)\citenamefont
  {Ricotti}, \citenamefont {Ostriker},\ and\ \citenamefont
  {Mack}}]{Ricotti:2007au}%
  \BibitemOpen
  \bibfield  {author} {\bibinfo {author} {\bibfnamefont {M.}~\bibnamefont
  {Ricotti}}, \bibinfo {author} {\bibfnamefont {J.~P.}\ \bibnamefont
  {Ostriker}}, \ and\ \bibinfo {author} {\bibfnamefont {K.~J.}\ \bibnamefont
  {Mack}},\ }\href {\doibase 10.1086/587831} {\bibfield  {journal} {\bibinfo
  {journal} {Astrophys. J.}\ }\textbf {\bibinfo {volume} {680}},\ \bibinfo
  {pages} {829} (\bibinfo {year} {2008})},\ \Eprint
  {http://arxiv.org/abs/0709.0524} {arXiv:0709.0524 [astro-ph]} \BibitemShut
  {NoStop}%
\bibitem [{\citenamefont {Horowitz}(2016)}]{Horowitz:2016lib}%
  \BibitemOpen
  \bibfield  {author} {\bibinfo {author} {\bibfnamefont {B.}~\bibnamefont
  {Horowitz}},\ }\href@noop {} {\  (\bibinfo {year} {2016})},\ \Eprint
  {http://arxiv.org/abs/1612.07264} {arXiv:1612.07264 [astro-ph.CO]}
  \BibitemShut {NoStop}%
\bibitem [{\citenamefont {Ali-Ha\"\i{}moud}\ and\ \citenamefont
  {Kamionkowski}(2017)}]{Ali-Haimoud:2016mbv}%
  \BibitemOpen
  \bibfield  {author} {\bibinfo {author} {\bibfnamefont {Y.}~\bibnamefont
  {Ali-Ha\"\i{}moud}}\ and\ \bibinfo {author} {\bibfnamefont {M.}~\bibnamefont
  {Kamionkowski}},\ }\href {\doibase 10.1103/PhysRevD.95.043534} {\bibfield
  {journal} {\bibinfo  {journal} {Phys. Rev. D}\ }\textbf {\bibinfo {volume}
  {95}},\ \bibinfo {pages} {043534} (\bibinfo {year} {2017})},\ \Eprint
  {http://arxiv.org/abs/1612.05644} {arXiv:1612.05644 [astro-ph.CO]}
  \BibitemShut {NoStop}%
\bibitem [{\citenamefont {Poulin}\ \emph {et~al.}(2017)\citenamefont {Poulin},
  \citenamefont {Serpico}, \citenamefont {Calore}, \citenamefont {Clesse},\
  and\ \citenamefont {Kohri}}]{Poulin:2017bwe}%
  \BibitemOpen
  \bibfield  {author} {\bibinfo {author} {\bibfnamefont {V.}~\bibnamefont
  {Poulin}}, \bibinfo {author} {\bibfnamefont {P.~D.}\ \bibnamefont {Serpico}},
  \bibinfo {author} {\bibfnamefont {F.}~\bibnamefont {Calore}}, \bibinfo
  {author} {\bibfnamefont {S.}~\bibnamefont {Clesse}}, \ and\ \bibinfo {author}
  {\bibfnamefont {K.}~\bibnamefont {Kohri}},\ }\href {\doibase
  10.1103/PhysRevD.96.083524} {\bibfield  {journal} {\bibinfo  {journal} {Phys.
  Rev.}\ }\textbf {\bibinfo {volume} {D96}},\ \bibinfo {pages} {083524}
  (\bibinfo {year} {2017})},\ \Eprint {http://arxiv.org/abs/1707.04206}
  {arXiv:1707.04206 [astro-ph.CO]} \BibitemShut {NoStop}%
\bibitem [{\citenamefont {Hektor}\ \emph {et~al.}(2018)\citenamefont {Hektor},
  \citenamefont {{{H\"utsi}}}, \citenamefont {Marzola}, \citenamefont {Raidal},
  \citenamefont {Vaskonen},\ and\ \citenamefont
  {{{Veerm\"ae}}}}]{Hektor:2018qqw}%
  \BibitemOpen
  \bibfield  {author} {\bibinfo {author} {\bibfnamefont {A.}~\bibnamefont
  {Hektor}}, \bibinfo {author} {\bibfnamefont {G.}~\bibnamefont {{{H\"utsi}}}},
  \bibinfo {author} {\bibfnamefont {L.}~\bibnamefont {Marzola}}, \bibinfo
  {author} {\bibfnamefont {M.}~\bibnamefont {Raidal}}, \bibinfo {author}
  {\bibfnamefont {V.}~\bibnamefont {Vaskonen}}, \ and\ \bibinfo {author}
  {\bibfnamefont {H.}~\bibnamefont {{{Veerm\"ae}}}},\ }\href {\doibase
  10.1103/PhysRevD.98.023503} {\bibfield  {journal} {\bibinfo  {journal} {Phys.
  Rev.}\ }\textbf {\bibinfo {volume} {D98}},\ \bibinfo {pages} {023503}
  (\bibinfo {year} {2018})},\ \Eprint {http://arxiv.org/abs/1803.09697}
  {arXiv:1803.09697 [astro-ph.CO]} \BibitemShut {NoStop}%
\bibitem [{\citenamefont {Serpico}\ \emph {et~al.}(2020)\citenamefont
  {Serpico}, \citenamefont {Poulin}, \citenamefont {Inman},\ and\ \citenamefont
  {Kohri}}]{Serpico:2020ehh}%
  \BibitemOpen
  \bibfield  {author} {\bibinfo {author} {\bibfnamefont {P.~D.}\ \bibnamefont
  {Serpico}}, \bibinfo {author} {\bibfnamefont {V.}~\bibnamefont {Poulin}},
  \bibinfo {author} {\bibfnamefont {D.}~\bibnamefont {Inman}}, \ and\ \bibinfo
  {author} {\bibfnamefont {K.}~\bibnamefont {Kohri}},\ }\href {\doibase
  10.1103/PhysRevResearch.2.023204} {\bibfield  {journal} {\bibinfo  {journal}
  {Phys. Rev. Res.}\ }\textbf {\bibinfo {volume} {2}},\ \bibinfo {pages}
  {023204} (\bibinfo {year} {2020})},\ \Eprint
  {http://arxiv.org/abs/2002.10771} {arXiv:2002.10771 [astro-ph.CO]}
  \BibitemShut {NoStop}%
\bibitem [{\citenamefont {De~Luca}\ \emph
  {et~al.}(2020{\natexlab{c}})\citenamefont {De~Luca}, \citenamefont
  {Franciolini}, \citenamefont {Pani},\ and\ \citenamefont
  {Riotto}}]{DeLuca:2020fpg}%
  \BibitemOpen
  \bibfield  {author} {\bibinfo {author} {\bibfnamefont {V.}~\bibnamefont
  {De~Luca}}, \bibinfo {author} {\bibfnamefont {G.}~\bibnamefont
  {Franciolini}}, \bibinfo {author} {\bibfnamefont {P.}~\bibnamefont {Pani}}, \
  and\ \bibinfo {author} {\bibfnamefont {A.}~\bibnamefont {Riotto}},\ }\href
  {\doibase 10.1103/PhysRevD.102.043505} {\bibfield  {journal} {\bibinfo
  {journal} {Phys. Rev. D}\ }\textbf {\bibinfo {volume} {102}},\ \bibinfo
  {pages} {043505} (\bibinfo {year} {2020}{\natexlab{c}})},\ \Eprint
  {http://arxiv.org/abs/2003.12589} {arXiv:2003.12589 [astro-ph.CO]}
  \BibitemShut {NoStop}%
\bibitem [{\citenamefont {{Thakurta}}(1981)}]{1981InJPh..55..304T}%
  \BibitemOpen
  \bibfield  {author} {\bibinfo {author} {\bibfnamefont {S.~N.~G.}\
  \bibnamefont {{Thakurta}}},\ }\href@noop {} {\bibfield  {journal} {\bibinfo
  {journal} {Indian Journal of Physics}\ }\textbf {\bibinfo {volume} {55B}},\
  \bibinfo {pages} {304} (\bibinfo {year} {1981})}\BibitemShut {NoStop}%
\bibitem [{\citenamefont {Boehm}\ \emph
  {et~al.}(2021{\natexlab{a}})\citenamefont {Boehm}, \citenamefont
  {Kobakhidze}, \citenamefont {O'hare}, \citenamefont {Picker},\ and\
  \citenamefont {Sakellariadou}}]{Boehm:2020jwd}%
  \BibitemOpen
  \bibfield  {author} {\bibinfo {author} {\bibfnamefont {C.}~\bibnamefont
  {Boehm}}, \bibinfo {author} {\bibfnamefont {A.}~\bibnamefont {Kobakhidze}},
  \bibinfo {author} {\bibfnamefont {C.~A.~J.}\ \bibnamefont {O'hare}}, \bibinfo
  {author} {\bibfnamefont {Z.~S.~C.}\ \bibnamefont {Picker}}, \ and\ \bibinfo
  {author} {\bibfnamefont {M.}~\bibnamefont {Sakellariadou}},\ }\href {\doibase
  10.1088/1475-7516/2021/03/078} {\bibfield  {journal} {\bibinfo  {journal}
  {JCAP}\ }\textbf {\bibinfo {volume} {03}},\ \bibinfo {pages} {078} (\bibinfo
  {year} {2021}{\natexlab{a}})},\ \Eprint {http://arxiv.org/abs/2008.10743}
  {arXiv:2008.10743 [astro-ph.CO]} \BibitemShut {NoStop}%
\bibitem [{\citenamefont {Picker}(2021)}]{Picker:2021jxl}%
  \BibitemOpen
  \bibfield  {author} {\bibinfo {author} {\bibfnamefont {Z.~S.~C.}\
  \bibnamefont {Picker}},\ }\href@noop {} {\  (\bibinfo {year} {2021})},\
  \Eprint {http://arxiv.org/abs/2103.02815} {arXiv:2103.02815 [astro-ph.CO]}
  \BibitemShut {NoStop}%
\bibitem [{\citenamefont {Faraoni}(2018)}]{Faraoni:2018xwo}%
  \BibitemOpen
  \bibfield  {author} {\bibinfo {author} {\bibfnamefont {V.}~\bibnamefont
  {Faraoni}},\ }\href {\doibase 10.3390/universe4100109} {\bibfield  {journal}
  {\bibinfo  {journal} {Universe}\ }\textbf {\bibinfo {volume} {4}},\ \bibinfo
  {pages} {109} (\bibinfo {year} {2018})},\ \Eprint
  {http://arxiv.org/abs/1810.04667} {arXiv:1810.04667 [gr-qc]} \BibitemShut
  {NoStop}%
\bibitem [{\citenamefont {Hayward}(1996)}]{Hayward:1994bu}%
  \BibitemOpen
  \bibfield  {author} {\bibinfo {author} {\bibfnamefont {S.~A.}\ \bibnamefont
  {Hayward}},\ }\href {\doibase 10.1103/PhysRevD.53.1938} {\bibfield  {journal}
  {\bibinfo  {journal} {Phys. Rev. D}\ }\textbf {\bibinfo {volume} {53}},\
  \bibinfo {pages} {1938} (\bibinfo {year} {1996})},\ \Eprint
  {http://arxiv.org/abs/gr-qc/9408002} {arXiv:gr-qc/9408002} \BibitemShut
  {NoStop}%
\bibitem [{\citenamefont {Carrera}\ and\ \citenamefont
  {Giulini}(2010)}]{Carrera:2009ve}%
  \BibitemOpen
  \bibfield  {author} {\bibinfo {author} {\bibfnamefont {M.}~\bibnamefont
  {Carrera}}\ and\ \bibinfo {author} {\bibfnamefont {D.}~\bibnamefont
  {Giulini}},\ }\href {\doibase 10.1103/PhysRevD.81.043521} {\bibfield
  {journal} {\bibinfo  {journal} {Phys. Rev. D}\ }\textbf {\bibinfo {volume}
  {81}},\ \bibinfo {pages} {043521} (\bibinfo {year} {2010})},\ \Eprint
  {http://arxiv.org/abs/0908.3101} {arXiv:0908.3101 [gr-qc]} \BibitemShut
  {NoStop}%
\bibitem [{\citenamefont {Mello}\ \emph {et~al.}(2017)\citenamefont {Mello},
  \citenamefont {Maciel},\ and\ \citenamefont {Zanchin}}]{Mello:2016irl}%
  \BibitemOpen
  \bibfield  {author} {\bibinfo {author} {\bibfnamefont {M.~M.~C.}\
  \bibnamefont {Mello}}, \bibinfo {author} {\bibfnamefont {A.}~\bibnamefont
  {Maciel}}, \ and\ \bibinfo {author} {\bibfnamefont {V.~T.}\ \bibnamefont
  {Zanchin}},\ }\href {\doibase 10.1103/PhysRevD.95.084031} {\bibfield
  {journal} {\bibinfo  {journal} {Phys. Rev. D}\ }\textbf {\bibinfo {volume}
  {95}},\ \bibinfo {pages} {084031} (\bibinfo {year} {2017})},\ \Eprint
  {http://arxiv.org/abs/1611.05077} {arXiv:1611.05077 [gr-qc]} \BibitemShut
  {NoStop}%
\bibitem [{\citenamefont {H\"utsi}\ \emph {et~al.}(2019)\citenamefont
  {H\"utsi}, \citenamefont {Raidal},\ and\ \citenamefont
  {Veerm\"ae}}]{Hutsi:2019hlw}%
  \BibitemOpen
  \bibfield  {author} {\bibinfo {author} {\bibfnamefont {G.}~\bibnamefont
  {H\"utsi}}, \bibinfo {author} {\bibfnamefont {M.}~\bibnamefont {Raidal}}, \
  and\ \bibinfo {author} {\bibfnamefont {H.}~\bibnamefont {Veerm\"ae}},\ }\href
  {\doibase 10.1103/PhysRevD.100.083016} {\bibfield  {journal} {\bibinfo
  {journal} {Phys. Rev. D}\ }\textbf {\bibinfo {volume} {100}},\ \bibinfo
  {pages} {083016} (\bibinfo {year} {2019})},\ \Eprint
  {http://arxiv.org/abs/1907.06533} {arXiv:1907.06533 [astro-ph.CO]}
  \BibitemShut {NoStop}%
\bibitem [{\citenamefont {Inman}\ and\ \citenamefont
  {Ali-Haïmoud}(2019)}]{Inman:2019wvr}%
  \BibitemOpen
  \bibfield  {author} {\bibinfo {author} {\bibfnamefont {D.}~\bibnamefont
  {Inman}}\ and\ \bibinfo {author} {\bibfnamefont {Y.}~\bibnamefont
  {Ali-Haïmoud}},\ }\href {\doibase 10.1103/PhysRevD.100.083528} {\bibfield
  {journal} {\bibinfo  {journal} {Phys. Rev. D}\ }\textbf {\bibinfo {volume}
  {100}},\ \bibinfo {pages} {083528} (\bibinfo {year} {2019})},\ \Eprint
  {http://arxiv.org/abs/1907.08129} {arXiv:1907.08129 [astro-ph.CO]}
  \BibitemShut {NoStop}%
\bibitem [{\citenamefont {De~Luca}\ \emph
  {et~al.}(2020{\natexlab{d}})\citenamefont {De~Luca}, \citenamefont
  {Desjacques}, \citenamefont {Franciolini},\ and\ \citenamefont
  {Riotto}}]{DeLuca:2020jug}%
  \BibitemOpen
  \bibfield  {author} {\bibinfo {author} {\bibfnamefont {V.}~\bibnamefont
  {De~Luca}}, \bibinfo {author} {\bibfnamefont {V.}~\bibnamefont {Desjacques}},
  \bibinfo {author} {\bibfnamefont {G.}~\bibnamefont {Franciolini}}, \ and\
  \bibinfo {author} {\bibfnamefont {A.}~\bibnamefont {Riotto}},\ }\href
  {\doibase 10.1088/1475-7516/2020/11/028} {\bibfield  {journal} {\bibinfo
  {journal} {JCAP}\ }\textbf {\bibinfo {volume} {11}},\ \bibinfo {pages} {028}
  (\bibinfo {year} {2020}{\natexlab{d}})},\ \Eprint
  {http://arxiv.org/abs/2009.04731} {arXiv:2009.04731 [astro-ph.CO]}
  \BibitemShut {NoStop}%
\bibitem [{\citenamefont {Tseliakhovich}\ and\ \citenamefont
  {Hirata}(2010)}]{Tseliakhovich:2010bj}%
  \BibitemOpen
  \bibfield  {author} {\bibinfo {author} {\bibfnamefont {D.}~\bibnamefont
  {Tseliakhovich}}\ and\ \bibinfo {author} {\bibfnamefont {C.}~\bibnamefont
  {Hirata}},\ }\href {\doibase 10.1103/PhysRevD.82.083520} {\bibfield
  {journal} {\bibinfo  {journal} {Phys. Rev. D}\ }\textbf {\bibinfo {volume}
  {82}},\ \bibinfo {pages} {083520} (\bibinfo {year} {2010})},\ \Eprint
  {http://arxiv.org/abs/1005.2416} {arXiv:1005.2416 [astro-ph.CO]} \BibitemShut
  {NoStop}%
\bibitem [{\citenamefont {Clesse}\ and\ \citenamefont
  {García-Bellido}(2017)}]{Clesse:2016vqa}%
  \BibitemOpen
  \bibfield  {author} {\bibinfo {author} {\bibfnamefont {S.}~\bibnamefont
  {Clesse}}\ and\ \bibinfo {author} {\bibfnamefont {J.}~\bibnamefont
  {García-Bellido}},\ }\href {\doibase 10.1016/j.dark.2016.10.002} {\bibfield
  {journal} {\bibinfo  {journal} {Phys. Dark Univ.}\ }\textbf {\bibinfo
  {volume} {15}},\ \bibinfo {pages} {142} (\bibinfo {year} {2017})},\ \Eprint
  {http://arxiv.org/abs/1603.05234} {arXiv:1603.05234 [astro-ph.CO]}
  \BibitemShut {NoStop}%
\bibitem [{\citenamefont {Ballesteros}\ \emph {et~al.}(2018)\citenamefont
  {Ballesteros}, \citenamefont {Serpico},\ and\ \citenamefont
  {Taoso}}]{Ballesteros:2018swv}%
  \BibitemOpen
  \bibfield  {author} {\bibinfo {author} {\bibfnamefont {G.}~\bibnamefont
  {Ballesteros}}, \bibinfo {author} {\bibfnamefont {P.~D.}\ \bibnamefont
  {Serpico}}, \ and\ \bibinfo {author} {\bibfnamefont {M.}~\bibnamefont
  {Taoso}},\ }\href {\doibase 10.1088/1475-7516/2018/10/043} {\bibfield
  {journal} {\bibinfo  {journal} {JCAP}\ }\textbf {\bibinfo {volume} {10}},\
  \bibinfo {pages} {043} (\bibinfo {year} {2018})},\ \Eprint
  {http://arxiv.org/abs/1807.02084} {arXiv:1807.02084 [astro-ph.CO]}
  \BibitemShut {NoStop}%
\bibitem [{\citenamefont {Young}\ and\ \citenamefont
  {Byrnes}(2020)}]{Young:2019gfc}%
  \BibitemOpen
  \bibfield  {author} {\bibinfo {author} {\bibfnamefont {S.}~\bibnamefont
  {Young}}\ and\ \bibinfo {author} {\bibfnamefont {C.~T.}\ \bibnamefont
  {Byrnes}},\ }\href {\doibase 10.1088/1475-7516/2020/03/004} {\bibfield
  {journal} {\bibinfo  {journal} {JCAP}\ }\textbf {\bibinfo {volume} {03}},\
  \bibinfo {pages} {004} (\bibinfo {year} {2020})},\ \Eprint
  {http://arxiv.org/abs/1910.06077} {arXiv:1910.06077 [astro-ph.CO]}
  \BibitemShut {NoStop}%
\bibitem [{\citenamefont {Eroshenko}(2021)}]{Eroshenko:2021oeq}%
  \BibitemOpen
  \bibfield  {author} {\bibinfo {author} {\bibfnamefont {Y.}~\bibnamefont
  {Eroshenko}},\ }\href {\doibase 10.1016/j.dark.2021.100833} {\bibfield
  {journal} {\bibinfo  {journal} {Phys. Dark Univ.}\ }\textbf {\bibinfo
  {volume} {32}},\ \bibinfo {pages} {100833} (\bibinfo {year} {2021})},\
  \Eprint {http://arxiv.org/abs/2105.03704} {arXiv:2105.03704 [astro-ph.CO]}
  \BibitemShut {NoStop}%
\bibitem [{\citenamefont {De~Luca}\ \emph
  {et~al.}(2021{\natexlab{b}})\citenamefont {De~Luca}, \citenamefont
  {Franciolini},\ and\ \citenamefont {Riotto}}]{DeLuca:2021hcf}%
  \BibitemOpen
  \bibfield  {author} {\bibinfo {author} {\bibfnamefont {V.}~\bibnamefont
  {De~Luca}}, \bibinfo {author} {\bibfnamefont {G.}~\bibnamefont
  {Franciolini}}, \ and\ \bibinfo {author} {\bibfnamefont {A.}~\bibnamefont
  {Riotto}},\ }\href@noop {} {\  (\bibinfo {year} {2021}{\natexlab{b}})},\
  \Eprint {http://arxiv.org/abs/2103.16369} {arXiv:2103.16369 [astro-ph.CO]}
  \BibitemShut {NoStop}%
\bibitem [{\citenamefont {Adamek}\ \emph {et~al.}(2019)\citenamefont {Adamek},
  \citenamefont {Byrnes}, \citenamefont {Gosenca},\ and\ \citenamefont
  {Hotchkiss}}]{Adamek:2019gns}%
  \BibitemOpen
  \bibfield  {author} {\bibinfo {author} {\bibfnamefont {J.}~\bibnamefont
  {Adamek}}, \bibinfo {author} {\bibfnamefont {C.~T.}\ \bibnamefont {Byrnes}},
  \bibinfo {author} {\bibfnamefont {M.}~\bibnamefont {Gosenca}}, \ and\
  \bibinfo {author} {\bibfnamefont {S.}~\bibnamefont {Hotchkiss}},\ }\href
  {\doibase 10.1103/PhysRevD.100.023506} {\bibfield  {journal} {\bibinfo
  {journal} {Phys. Rev. D}\ }\textbf {\bibinfo {volume} {100}},\ \bibinfo
  {pages} {023506} (\bibinfo {year} {2019})},\ \Eprint
  {http://arxiv.org/abs/1901.08528} {arXiv:1901.08528 [astro-ph.CO]}
  \BibitemShut {NoStop}%
\bibitem [{\citenamefont {Farris}\ \emph {et~al.}(2014)\citenamefont {Farris},
  \citenamefont {Duffell}, \citenamefont {MacFadyen},\ and\ \citenamefont
  {Haiman}}]{Farris:2013uqa}%
  \BibitemOpen
  \bibfield  {author} {\bibinfo {author} {\bibfnamefont {B.~D.}\ \bibnamefont
  {Farris}}, \bibinfo {author} {\bibfnamefont {P.}~\bibnamefont {Duffell}},
  \bibinfo {author} {\bibfnamefont {A.~I.}\ \bibnamefont {MacFadyen}}, \ and\
  \bibinfo {author} {\bibfnamefont {Z.}~\bibnamefont {Haiman}},\ }\href
  {\doibase 10.1088/0004-637X/783/2/134} {\bibfield  {journal} {\bibinfo
  {journal} {Astrophys. J.}\ }\textbf {\bibinfo {volume} {783}},\ \bibinfo
  {pages} {134} (\bibinfo {year} {2014})},\ \Eprint
  {http://arxiv.org/abs/1310.0492} {arXiv:1310.0492 [astro-ph.HE]} \BibitemShut
  {NoStop}%
\bibitem [{\citenamefont {Duez}\ and\ \citenamefont
  {Zlochower}(2019)}]{Duez:2018jaf}%
  \BibitemOpen
  \bibfield  {author} {\bibinfo {author} {\bibfnamefont {M.~D.}\ \bibnamefont
  {Duez}}\ and\ \bibinfo {author} {\bibfnamefont {Y.}~\bibnamefont
  {Zlochower}},\ }\href {\doibase 10.1088/1361-6633/aadb16} {\bibfield
  {journal} {\bibinfo  {journal} {Rept. Prog. Phys.}\ }\textbf {\bibinfo
  {volume} {82}},\ \bibinfo {pages} {016902} (\bibinfo {year} {2019})},\
  \Eprint {http://arxiv.org/abs/1808.06011} {arXiv:1808.06011 [gr-qc]}
  \BibitemShut {NoStop}%
\bibitem [{\citenamefont {Mu\~noz}\ \emph {et~al.}(2019)\citenamefont
  {Mu\~noz}, \citenamefont {Miranda},\ and\ \citenamefont
  {Lai}}]{Munoz:2018tnj}%
  \BibitemOpen
  \bibfield  {author} {\bibinfo {author} {\bibfnamefont {D.~J.}\ \bibnamefont
  {Mu\~noz}}, \bibinfo {author} {\bibfnamefont {R.}~\bibnamefont {Miranda}}, \
  and\ \bibinfo {author} {\bibfnamefont {D.}~\bibnamefont {Lai}},\ }\href
  {\doibase 10.3847/1538-4357/aaf867} {\bibfield  {journal} {\bibinfo
  {journal} {Astrophys. J.}\ }\textbf {\bibinfo {volume} {871}},\ \bibinfo
  {pages} {84} (\bibinfo {year} {2019})},\ \Eprint
  {http://arxiv.org/abs/1810.04676} {arXiv:1810.04676 [astro-ph.HE]}
  \BibitemShut {NoStop}%
\bibitem [{\citenamefont {McVittie}(1933)}]{10.1093/mnras/93.5.325}%
  \BibitemOpen
  \bibfield  {author} {\bibinfo {author} {\bibfnamefont {G.~C.}\ \bibnamefont
  {McVittie}},\ }\href {\doibase 10.1093/mnras/93.5.325} {\bibfield  {journal}
  {\bibinfo  {journal} {Monthly Notices of the Royal Astronomical Society}\
  }\textbf {\bibinfo {volume} {93}},\ \bibinfo {pages} {325} (\bibinfo {year}
  {1933})},\ \Eprint
  {http://arxiv.org/abs/https://academic.oup.com/mnras/article-pdf/93/5/325/2793517/mnras93-0325.pdf}
  {https://academic.oup.com/mnras/article-pdf/93/5/325/2793517/mnras93-0325.pdf}
  \BibitemShut {NoStop}%
\bibitem [{\citenamefont {Faraoni}\ and\ \citenamefont
  {Jacques}(2007)}]{Faraoni:2007es}%
  \BibitemOpen
  \bibfield  {author} {\bibinfo {author} {\bibfnamefont {V.}~\bibnamefont
  {Faraoni}}\ and\ \bibinfo {author} {\bibfnamefont {A.}~\bibnamefont
  {Jacques}},\ }\href {\doibase 10.1103/PhysRevD.76.063510} {\bibfield
  {journal} {\bibinfo  {journal} {Phys. Rev. D}\ }\textbf {\bibinfo {volume}
  {76}},\ \bibinfo {pages} {063510} (\bibinfo {year} {2007})},\ \Eprint
  {http://arxiv.org/abs/0707.1350} {arXiv:0707.1350 [gr-qc]} \BibitemShut
  {NoStop}%
\bibitem [{\citenamefont {Faraoni}\ \emph {et~al.}(2009)\citenamefont
  {Faraoni}, \citenamefont {Gao}, \citenamefont {Chen},\ and\ \citenamefont
  {Shen}}]{Faraoni:2008tx}%
  \BibitemOpen
  \bibfield  {author} {\bibinfo {author} {\bibfnamefont {V.}~\bibnamefont
  {Faraoni}}, \bibinfo {author} {\bibfnamefont {C.}~\bibnamefont {Gao}},
  \bibinfo {author} {\bibfnamefont {X.}~\bibnamefont {Chen}}, \ and\ \bibinfo
  {author} {\bibfnamefont {Y.-G.}\ \bibnamefont {Shen}},\ }\href {\doibase
  10.1016/j.physletb.2008.11.067} {\bibfield  {journal} {\bibinfo  {journal}
  {Phys. Lett. B}\ }\textbf {\bibinfo {volume} {671}},\ \bibinfo {pages} {7}
  (\bibinfo {year} {2009})},\ \Eprint {http://arxiv.org/abs/0811.4667}
  {arXiv:0811.4667 [gr-qc]} \BibitemShut {NoStop}%
\bibitem [{\citenamefont {Bondi}(1952)}]{10.1093/mnras/112.2.195}%
  \BibitemOpen
  \bibfield  {author} {\bibinfo {author} {\bibfnamefont {H.}~\bibnamefont
  {Bondi}},\ }\href {\doibase 10.1093/mnras/112.2.195} {\bibfield  {journal}
  {\bibinfo  {journal} {Monthly Notices of the Royal Astronomical Society}\
  }\textbf {\bibinfo {volume} {112}},\ \bibinfo {pages} {195} (\bibinfo {year}
  {1952})},\ \Eprint
  {http://arxiv.org/abs/https://academic.oup.com/mnras/article-pdf/112/2/195/9073555/mnras112-0195.pdf}
  {https://academic.oup.com/mnras/article-pdf/112/2/195/9073555/mnras112-0195.pdf}
  \BibitemShut {NoStop}%
\bibitem [{\citenamefont {{Zel'dovich}}\ and\ \citenamefont
  {{Novikov}}(1967)}]{1967SvA....10..602Z}%
  \BibitemOpen
  \bibfield  {author} {\bibinfo {author} {\bibfnamefont {Y.~B.}\ \bibnamefont
  {{Zel'dovich}}}\ and\ \bibinfo {author} {\bibfnamefont {I.~D.}\ \bibnamefont
  {{Novikov}}},\ }\href@noop {} {\bibfield  {journal} {\bibinfo  {journal}
  {Soviet Astronomy}\ }\textbf {\bibinfo {volume} {10}},\ \bibinfo {pages}
  {602} (\bibinfo {year} {1967})}\BibitemShut {NoStop}%
\bibitem [{\citenamefont {Carr}\ and\ \citenamefont
  {Hawking}(1974)}]{10.1093/mnras/168.2.399}%
  \BibitemOpen
  \bibfield  {author} {\bibinfo {author} {\bibfnamefont {B.~J.}\ \bibnamefont
  {Carr}}\ and\ \bibinfo {author} {\bibfnamefont {S.~W.}\ \bibnamefont
  {Hawking}},\ }\href {\doibase 10.1093/mnras/168.2.399} {\bibfield  {journal}
  {\bibinfo  {journal} {Monthly Notices of the Royal Astronomical Society}\
  }\textbf {\bibinfo {volume} {168}},\ \bibinfo {pages} {399} (\bibinfo {year}
  {1974})},\ \Eprint
  {http://arxiv.org/abs/https://academic.oup.com/mnras/article-pdf/168/2/399/8079885/mnras168-0399.pdf}
  {https://academic.oup.com/mnras/article-pdf/168/2/399/8079885/mnras168-0399.pdf}
  \BibitemShut {NoStop}%
\bibitem [{\citenamefont {Babichev}\ \emph {et~al.}(2005)\citenamefont
  {Babichev}, \citenamefont {Dokuchaev},\ and\ \citenamefont
  {Eroshenko}}]{Babichev:2005py}%
  \BibitemOpen
  \bibfield  {author} {\bibinfo {author} {\bibfnamefont {E.}~\bibnamefont
  {Babichev}}, \bibinfo {author} {\bibfnamefont {V.}~\bibnamefont {Dokuchaev}},
  \ and\ \bibinfo {author} {\bibfnamefont {Y.}~\bibnamefont {Eroshenko}},\
  }\href {\doibase 10.1134/1.1901765} {\bibfield  {journal} {\bibinfo
  {journal} {J. Exp. Theor. Phys.}\ }\textbf {\bibinfo {volume} {100}},\
  \bibinfo {pages} {528} (\bibinfo {year} {2005})},\ \Eprint
  {http://arxiv.org/abs/astro-ph/0505618} {arXiv:astro-ph/0505618} \BibitemShut
  {NoStop}%
\bibitem [{\citenamefont {Babichev}\ \emph {et~al.}(2018)\citenamefont
  {Babichev}, \citenamefont {Dokuchaev},\ and\ \citenamefont
  {Eroshenko}}]{Babichev:2018ubo}%
  \BibitemOpen
  \bibfield  {author} {\bibinfo {author} {\bibfnamefont {E.}~\bibnamefont
  {Babichev}}, \bibinfo {author} {\bibfnamefont {V.}~\bibnamefont {Dokuchaev}},
  \ and\ \bibinfo {author} {\bibfnamefont {Y.~N.}\ \bibnamefont {Eroshenko}},\
  }\href {\doibase 10.1134/S1063773718090013} {\bibfield  {journal} {\bibinfo
  {journal} {Astron. Lett.}\ }\textbf {\bibinfo {volume} {44}},\ \bibinfo
  {pages} {491} (\bibinfo {year} {2018})},\ \Eprint
  {http://arxiv.org/abs/1811.07189} {arXiv:1811.07189 [gr-qc]} \BibitemShut
  {NoStop}%
\bibitem [{\citenamefont {Carr}\ \emph {et~al.}(2010)\citenamefont {Carr},
  \citenamefont {Harada},\ and\ \citenamefont {Maeda}}]{Carr:2010wk}%
  \BibitemOpen
  \bibfield  {author} {\bibinfo {author} {\bibfnamefont {B.}~\bibnamefont
  {Carr}}, \bibinfo {author} {\bibfnamefont {T.}~\bibnamefont {Harada}}, \ and\
  \bibinfo {author} {\bibfnamefont {H.}~\bibnamefont {Maeda}},\ }\href
  {\doibase 10.1088/0264-9381/27/18/183101} {\bibfield  {journal} {\bibinfo
  {journal} {Class. Quant. Grav.}\ }\textbf {\bibinfo {volume} {27}},\ \bibinfo
  {pages} {183101} (\bibinfo {year} {2010})},\ \Eprint
  {http://arxiv.org/abs/1003.3324} {arXiv:1003.3324 [gr-qc]} \BibitemShut
  {NoStop}%
\bibitem [{\citenamefont {Boehm}\ \emph
  {et~al.}(2021{\natexlab{b}})\citenamefont {Boehm}, \citenamefont
  {Kobakhidze}, \citenamefont {O'Hare}, \citenamefont {Picker},\ and\
  \citenamefont {Sakellariadou}}]{Boehm:2021kzq}%
  \BibitemOpen
  \bibfield  {author} {\bibinfo {author} {\bibfnamefont {C.}~\bibnamefont
  {Boehm}}, \bibinfo {author} {\bibfnamefont {A.}~\bibnamefont {Kobakhidze}},
  \bibinfo {author} {\bibfnamefont {C.~A.~J.}\ \bibnamefont {O'Hare}}, \bibinfo
  {author} {\bibfnamefont {Z.~S.~C.}\ \bibnamefont {Picker}}, \ and\ \bibinfo
  {author} {\bibfnamefont {M.}~\bibnamefont {Sakellariadou}},\ }\href@noop {}
  {\  (\bibinfo {year} {2021}{\natexlab{b}})},\ \Eprint
  {http://arxiv.org/abs/2105.14908} {arXiv:2105.14908 [astro-ph.CO]}
  \BibitemShut {NoStop}%
\bibitem [{\citenamefont {H\"utsi}\ \emph
  {et~al.}(2021{\natexlab{b}})\citenamefont {H\"utsi}, \citenamefont
  {Koivisto}, \citenamefont {Raidal}, \citenamefont {Vaskonen},\ and\
  \citenamefont {Veerm\"ae}}]{Hutsi:2021vha}%
  \BibitemOpen
  \bibfield  {author} {\bibinfo {author} {\bibfnamefont {G.}~\bibnamefont
  {H\"utsi}}, \bibinfo {author} {\bibfnamefont {T.}~\bibnamefont {Koivisto}},
  \bibinfo {author} {\bibfnamefont {M.}~\bibnamefont {Raidal}}, \bibinfo
  {author} {\bibfnamefont {V.}~\bibnamefont {Vaskonen}}, \ and\ \bibinfo
  {author} {\bibfnamefont {H.}~\bibnamefont {Veerm\"ae}},\ }\href@noop {} {\
  (\bibinfo {year} {2021}{\natexlab{b}})},\ \Eprint
  {http://arxiv.org/abs/2106.02007} {arXiv:2106.02007 [astro-ph.CO]}
  \BibitemShut {NoStop}%
\bibitem [{\citenamefont {Harada}\ \emph {et~al.}(2021)\citenamefont {Harada},
  \citenamefont {Maeda},\ and\ \citenamefont {Sato}}]{Harada:2021xze}%
  \BibitemOpen
  \bibfield  {author} {\bibinfo {author} {\bibfnamefont {T.}~\bibnamefont
  {Harada}}, \bibinfo {author} {\bibfnamefont {H.}~\bibnamefont {Maeda}}, \
  and\ \bibinfo {author} {\bibfnamefont {T.}~\bibnamefont {Sato}},\ }\href@noop
  {} {\  (\bibinfo {year} {2021})},\ \Eprint {http://arxiv.org/abs/2106.06651}
  {arXiv:2106.06651 [gr-qc]} \BibitemShut {NoStop}%
\bibitem [{\citenamefont {Peters}(1964)}]{Peters:1964zz}%
  \BibitemOpen
  \bibfield  {author} {\bibinfo {author} {\bibfnamefont {P.~C.}\ \bibnamefont
  {Peters}},\ }\href {\doibase 10.1103/PhysRev.136.B1224} {\bibfield  {journal}
  {\bibinfo  {journal} {Phys. Rev.}\ }\textbf {\bibinfo {volume} {136}},\
  \bibinfo {pages} {B1224} (\bibinfo {year} {1964})}\BibitemShut {NoStop}%
\end{thebibliography}%
\end{document}